\documentclass{article}

\usepackage{latexsym}
\usepackage{epsfig}
\usepackage{citesort}
\usepackage{enumerate}

\begin{document} 


\newcommand{\smrfig}[2]{
  \begin{figure}[th]
  \begin{center}
  \input{#1.pstex_t}
  \caption{#2}
  \label{fig:#1}
  \end{center}
  \end{figure}
}

\renewcommand{\deg}{\ensuremath{^\circ}}
\newcommand{\calc}{\ensuremath{\mathcal{C}}}
\newcommand{\tmean}{\ensuremath{\bar{\theta}}}

\newtheorem{theorem}{Theorem}
\newtheorem{corollary}[theorem]{Corollary}
\newtheorem{lemma}[theorem]{Lemma}
\newtheorem{remark}[theorem]{Remark}

\newenvironment{proof}{\noindent\textbf{Proof:}}{\hfill $\Box$ \medskip}
\newenvironment{smalldef}{\medskip\noindent\textbf{Definition:}}{\medskip}

\newcommand{\proofitem}[1]{\smallskip\noindent\textbf{#1}}

\title{On Reconfiguring Tree Linkages:\\ Trees can Lock}

\setcounter{footnote}{-1}

\author{
Therese Biedl\thanks{%
  Research partially supported by 
  \textit{Fonds pour la Formation de Chercheurs et l'Aide \`a la
  Recherche}, Qu\'ebec (SW,SR), the
  \textit{Natural Sciences and Engineering Research Council}, Canada
  (AL,GT,SW,ED) and the
  \textit{National Science Foundation}, USA (JO,IS).
  }\thanks{
  University of Waterloo, Waterloo, Canada.
  \{biedl, eddemaine, mldemaine, alubiw\}@uwaterloo.ca.
  }
\hspace{3mm}
Erik Demaine\footnotemark[1]
\hspace{3mm}
Martin Demaine\footnotemark[1]
\\
Sylvain Lazard\thanks{
  INRIA Lorraine, France.
  lazard@loria.fr.
  }
\hspace{3mm}
Anna Lubiw\footnotemark[1]
\hspace{3mm}
Joseph O'Rourke\thanks{
  Smith College, Northampton, USA.
  \{orourke, streinu\}@cs.smith.edu.
  }
\hspace{3mm}
Steve Robbins\thanks{
  McGill University, Montreal, Canada.
  \{stever, godfried, sue\}@cgm.cs.mcgill.ca.
}\\
Ileana Streinu\footnotemark[3]
\hspace{3mm}
Godfried Toussaint\footnotemark[4]
\hspace{3mm}
Sue Whitesides\footnotemark[4]
}

\maketitle

\begin{abstract}
  It has recently been shown that any simple (i.e. nonintersecting)
  polygonal chain in the plane can be reconfigured to lie on a
  straight line, and any simple polygon can be reconfigured to be
  convex.  This result cannot be extended to tree linkages: we show
  that there are trees with two simple configurations that are not
  connected by a motion that preserves simplicity throughout the
  motion.  Indeed, we prove that an $N$-link tree can have
  $2^{\Omega(N)}$ equivalence classes of configurations.
\end{abstract}

\section{Introduction}

Consider a graph, each edge labelled with a positive number.  
Such a graph may be thought of as a collection of distance 
constraints between pairs of points in a Euclidean space.  A 
{\em realization} of such a graph maps each vertex to a point, 
also called a {\em joint}, and maps each edge 
to the closed line segment, called a {\em link},  connecting its
incident joints. 
The link length must equal the label of the
underlying graph edge.  
If a graph has one or more such realizations, we call it a
{\em linkage}. 

An embedding of a linkage in space is called a {\em configuration} of
the linkage if any pair of links whose underlying edges are incident
on a common vertex intersect only at the common joint and all other
pairs of links are disjoint.  
Some authors allow the term configuration to refer to objects that
self-intersect.  In contrast, we require all configurations to be
simple; i.e. non self-intersecting.
A {\em motion} of a linkage is a
continuous movement of its joints such that it remains in a valid
configuration at all times.  A natural question is whether a motion
exists between two given configurations of a linkage.

For a linkage in the plane whose underlying graph is a path, a related
question is whether it can always be \emph{straightened}; i.e.
whether it can be moved from any configuration to lie on a straight
line.  Similarly, we wonder whether a cycle linkage (polygon) can
always be \emph{convexified}; i.e. whether it can be moved to a
configuration that is a convex polygon.  If a linkage cannot be so
reconfigured, it is called \emph{locked}.  These questions have been
in the math community since the 1970's \cite{orourke00:column39} and
in the computational geometry community since 1991
\cite{lw-tpio-91,lw-rltm-92}, but first appeared in print in 1993 and
1995: \cite{lw93:rec-sim-pol} and \cite[p.~270]{k96:prob-low-dim-top}.
Initial computational geometry results focused on certain classes of
configurations such as ``visible'' chains \cite{visible-chains},
star-shaped polygons \cite{evlrsw98:con-star-shap-pol} and monotone
polygons \cite{monotone}.  Connelly, Demaine, and Rote have recently
proved that in the plane, no chain or polygon is locked
\cite{cdr00:str-pol}; Streinu \cite{s00:com-app} provides an
alternative proof.  In three dimensions, while a complete
characterization isn't known, there are configurations of open
polygonal chains and of polygons that can be straightened, or
convexified, respectively, and other configurations that can not be
\cite{chains3d}. In four or more dimensions, no chain or polygon is

Related linkage motion results in the
computational geometry literature
(e.g. \cite{hjw84:mov,hjw85,kk86:new,k92:mot,k97:reac,ksw96:fol,lw95:rec-euc,p96:rec-reac-chain,pw97:oen-fol,pw96:oen,pw96:rec-chain,sy96:des,w92:alg,k85:geom-inv-reac})
allow the links to cross or to pass through or over one another.  
In other words, the links represent distance
constraints between joints, not physical obstacles that must avoid each other.
There is also the very general algebraic approach to motion
planning of~\cite{c-crmp-87} and~\cite{ss83:oen-ii}, since the constraint
that the links not cross can be specified algebraically.  
For related work from a topological point of view, 
see~\cite{km95:mod-spac} and references therein;
again, this work allows links to cross.

There is a natural equivalence relation on the set of linkage
configurations: two configurations are equivalent if there is a motion
that takes the linkage from one configuration to the other.  The
Connelly-Demaine-Rote result states that chains in the plane have a single
equivalence class of configurations.  In this report we show that their
result cannot be generalized to tree linkages: trees can have many
configuration classes.

  \begin{figure}[th]
  \begin{center}
  \input{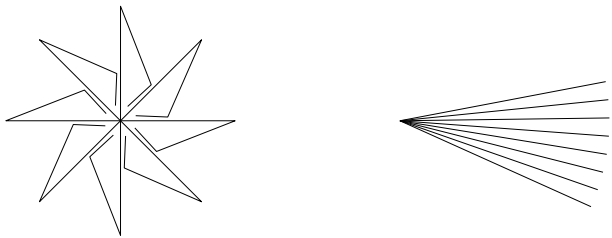}
  \caption{A tree linkage with disconnected configuration
space: no motion is possible from one configuration to the other.}
  \label{fig:inequivalent}
  \end{center}
  \end{figure}

This report establishes that a suitably-constructed tree linkage has
two configurations (pictured in Figure~\ref{fig:inequivalent}) for
which no motion between them is possible.  (This result also answers a
question posed in~\cite{dd97:com-ext-orig-bas}, arising from a paper
folding problem.)  As a corollary, we obtain the result that an
$N$-link tree linkage can exhibit $2^{\Omega(N)}$ equivalence classes
of configurations.

The rest of this report is organized as follows.
Section~\ref{sec:preliminaries} gives definitions and
the basic idea for constructing a locked
tree configuration,
Section~\ref{sec:construction} gives the construction itself, and
Sections~\ref{sec:restricted-config} and~\ref{sec:trees-can-lock} give
the correctness proof.  Section~\ref{sec:conclusion} concludes with
some open problems.

\section{Preliminaries}
\label{sec:preliminaries}

In this section, we introduce the technical definitions
used.  From now on, we often do not distinguish between 
vertices and edges in the underlying tree,  the corresponding 
joints and links in the tree linkage, and the points and 
line segments occupied by the joints in links in a particular 
configuration.  The context should make the meaning clear.  

\begin{figure}[th]
  \begin{center}
  \input{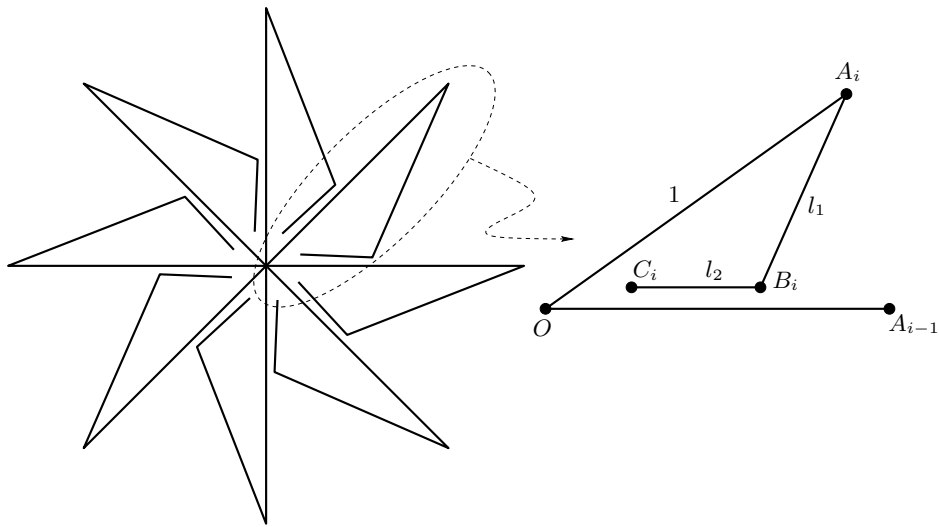}
  \caption{An example linkage with enlarged view of petal
$i$, showing the joint labels and the link lengths.}
  \label{fig:locked-tree}
  \end{center}
  \end{figure}

The trees considered in this report consist of $n$ {\em
petals}, each comprised of three links.  The initial configuration is
sketched in Figure~\ref{fig:locked-tree} and detailed as the report
unfolds.

The petals all meet at joint $O$; the other joints of petal $i$ are
labelled $A_i$, $B_i$, and $C_i$.  Designate the {\em petal angle} as
$\theta_i = \angle A_{i-1} O A_i$; let $\tmean = 2\pi / n$.  All index
arithmetic is taken mod $n$; all angles are measured in the interval
$[0,2\pi)$.  The link
lengths are as follows: $\|OA_i\| = 1$, $\|A_iB_i\| = l_1$, and
$\|B_iC_i\| = l_2$, where for
two points in the plane, $X$ and $Y$, we use $XY$
to designate the closed line segment between $X$ and $Y$, and $\|XY\|$
to denote its length.  
The values of $l_1$ and $l_2$ will be discussed in
Section~\ref{sec:construction}.

We often focus on
a single petal at a time, so the notation is simplified by suppressing
the petal index.  Joints of petal $i$ are referred to as $A$, $B$, and
$C$, joint $A_{i-1}$ is denoted $A'$, and the petal angle, i.e.,
$\angle A'OA$, as just $\theta$.  Let $L$ be the line through $OA'$, and
choose a reference frame with
$L$ oriented horizontally, $O$ to the left of $A'$.

\subsection{The Intuition}
\label{sec:intuition}

We first give a brief outline of the main argument, before delving
into the details.

  \begin{figure}[th]
  \begin{center}
  \input{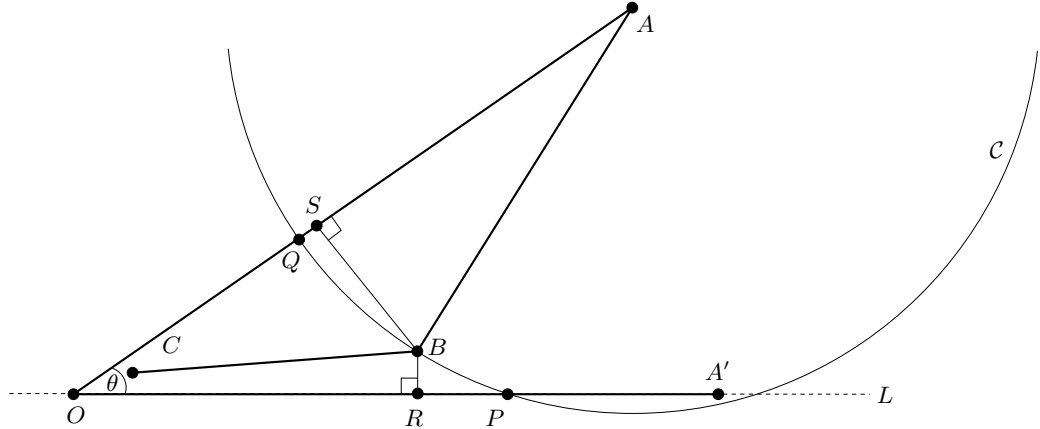}
  \caption{Illustrating the definition of {\em restricted
configuration}.}
  \label{fig:restricted-config}
  \end{center}
  \end{figure}

In the initial configuration, all petals are in a congruent
configuration, pictured in Figure~\ref{fig:restricted-config}, with
petal angle $\tmean = 2\pi/n$.  By choosing $l_1$ long enough, we
ensure that link $AB$ cannot swing out so as to straighten joint $A$.
Thus, the petal angle must be increased before joint $A$ can
straighten.  Because the petals all join at $O$, opening up one petal
necessitates squeezing the other petals.

Consider now how small we can squeeze a petal angle.  By choosing
$l_2$ long enough, we ensure that link $BC$ cannot swing over to fold
up against link $AB$.  In fact, joint $C$ is trapped inside
quadrilateral $\Box ORBS$.  We show later that the smallest petal
angle is obtained by moving joint $C$ to $O$, pictured in
Figure~\ref{fig:thetam}.  Lengthening $l_2$
increases this minimum petal angle, so by choosing $l_2$ long
enough, we can ensure that squeezing $n-1$ petals to their minimum
angle is still not enough to let the last petal open.
This, in essence, is why the tree is locked.  

From the above discussion, we note that $l_1$ and $l_2$ must be chosen
``long enough''.  On the other hand, they must also be ``short
enough'' that the configuration shown in
Figure~\ref{fig:restricted-config} is achievable.
Section~\ref{sec:construction} details all the constraints needed to
satisfy these two requirements.  Section~\ref{sec:restricted-config}
defines a ``restricted'' class of petal configurations (to which the
configuration in Figure~\ref{fig:restricted-config} belongs) and
proves that there is a non-zero minimum petal angle for this class of
configurations.  Finally, Section~\ref{sec:trees-can-lock} proves that
a linkage with parameters $n$, $l_1$, and $l_2$, satisfying all the
constraints of Section~\ref{sec:construction}, can be put into a 
initial configuration that is locked.

\section{Constructing a Locked Tree}
\label{sec:construction}

This section details the constraints on the parameters $n$, $l_1$, and
$l_2$ that are necessary to construct our locked linkage.
Lemma~\ref{lem:realizable}, at the end of this section,
demonstrates that the constraints are simultaneously satisfiable.
Refer to Figure~\ref{fig:restricted-config} throughout this section.

A three-link petal cannot be locked if the petal angle is greater than
or equal to $\pi/2$.  Thus we must have the initial petal angle
($\tmean = 2\pi/n$) strictly less than $\pi/2$, or
\begin{equation}
\label{ieq:n-lower}
        n > 4.
\end{equation}

Henceforth, assume $\theta < \pi/2$.  We want to have joint $B$ to the
left of the vertical line through $A$.  In order that link $AB$ fits,
we must have
\begin{equation}
\label{ieq:l1-upper}
        l_1 < 1.
\end{equation}

For any configuration of a petal, let $\calc$ be the circle centred at
$A$ with radius $l_1$.  Let $\beta$ be the petal angle at which \calc\
is tangent to $L$,
\begin{equation}
\label{eqn:define-beta}
        \beta = \arcsin l_1.
\end{equation}
The range for the principal value of arcsine is $[-\pi/2,\pi/2]$.
However, we know that $l_1$ must be positive, and less than 1
(Inequality~\ref{ieq:l1-upper}), so $\beta$ is actually in the range
$(0,\pi/2)$.

When $\theta < \beta$, circle \calc\ will properly intersect $L$. 
We want this to be true for the initial configuration, in
which all petal angles are $\tmean$, so we require
\begin{equation}
\label{ieq:tmean-upper}
        \tmean < \beta.
\end{equation}

Suppose \calc\ properly intersects $L$ and let $P$ be the leftmost
intersection point.  
Applying the cosine rule to $\triangle OAP$ 
and noting that $\|OA\| = 1$ and $\|AP\| = l_1$, yields
$l_1^2 = \|OP\|^2 + 1 - 2 \|OP\| \cos\theta$.
Since $l_1$, the radius of \calc, is strictly
less than $\|OA\| = 1$ by Inequality~\ref{ieq:l1-upper},
joint $O$ is outside the circle \calc.  From this, and
$\theta < \pi/2$, it follows that $P$ is to the right of $O$, and hence
$\|OP\|$ is the smaller root of the quadratic equation.
We define the function
\begin{equation}
\label{eqn:define-lP}
        l_P(\theta) 
                = \|OP\| 
                = \cos\theta - \sqrt{l_1^2 - \sin^2\theta}.
\end{equation}
This function is only defined on values of $\theta$ for which \calc\
intersects $L$; i.e., for $\theta$ in $[0,\beta]$. 
Differentiating~\ref{eqn:define-lP} shows
that $l_P$ is a strictly increasing function on this interval.

Furthermore, $P$ is to the left of the vertical line through $A$, and
hence $P \in OA'$, but $P$ is not at $A'$.  We know that $A$ is inside
the circle \calc, while $O$ is outside, so \calc\ intersects $OA$; let $Q$
be this intersection point.  Note that the small arc $PQ$ of \calc\
lies inside $\triangle A'OA$.

Suppose $\theta < \beta$, and $B$ is on the small arc $PQ$ of \calc.  Then $B$
is inside $\triangle A'OA$.  Since the triangle is isosceles, and the
angle $\theta$ is $< \pi/2$, $\triangle A'OA$ is acute.  This ensures
that there is a line passing through $B$ perpendicular to each edge of the
triangle.  Let $R \in OA'$ be such that $BR \perp OA'$ and $S \in OA$
be such that $BS \perp OA$.  We want to have joint $C$ inside
quadrilateral $\Box ORBS$.  This is feasible for the initial
configuration if we choose
\begin{equation}
\label{ieq:l2-upper}
        l_2 < l_P(\tmean),
\end{equation}
for then we may place $B$ near
$P$, and $C$ along $OB$.  At the same time, we wish to have $l_2$ long
enough that $C$ remains trapped in $\Box ORBS$.  This is ensured by
choosing
\begin{equation}
\label{ieq:l2-lower}
        l_2 > \sin\beta \cos\beta,
\end{equation}
as we show later in Section~\ref{sec:restricted-config}.

Define
\begin{equation}
\label{eqn:define-alpha}
        \alpha = (2\pi - \beta) / (n-1).
\end{equation}
Later, we will see that if a petal angle is opened past $\beta$, then
some other petal angle must be smaller than $\alpha$.  The proof that the
tree is locked then hinges on showing that there is a minimal petal
angle, which is greater than $\alpha$.

\begin{figure}[th]
  \begin{center}
  \input{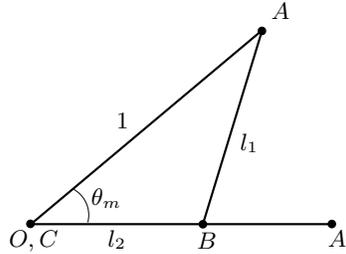}
  \caption{For calculating $\theta_m$.}
  \label{fig:thetam}
  \end{center}
  \end{figure}

To obtain a non-zero minimal petal angle, we require
\begin{equation}
\label{ieq:l1-plus-l2}
        l_1 + l_2 > 1,
\end{equation}
for otherwise, links $AB$ and $BC$ could fold flat against link $OA$,
and the petal angle could squeeze to zero.  
Indeed, we show later, in Lemma~\ref{lem:thetam-is-minimum}, that
the minimum possible petal angle is bounded from
below by the petal angle obtained in the non-simple configuration with
$C$ at $O$, and $B \in OA'$, pictured in Figure~\ref{fig:thetam}.
Define $\theta_m$ to be the resulting petal angle.  With $\triangle
OAB$, the cosine rule yields
\(
        l_1^2 = l_2^2 + 1 - 2l_2\cos\theta_m,
\)
or
\begin{equation}
\label{eqn:define-thetam}
       \theta_m = \arccos\left(\frac{1 - l_1^2 + l_2^2}{2 l_2}\right).
\end{equation}
We are using the principal value of arccosine, so
$\theta_m \in [0,\pi]$.

Finally, in order to prove Theorem~\ref{thm:main}, we assume that
$l_1$ and $l_2$ are such that
\begin{equation}
\label{ieq:alpha-upper}
        \alpha < \theta_m.
\end{equation}

We prove, in Appendix~\ref{app:realizable-proof}, the following
lemma which states that all the constraints of this section may be
simultaneously satisfied by an appropriate choice of $n$, $l_1$, and
$l_2$.  
For $n=5$, calculation shows that $l_1 = 0.9511$ and $l_2 = 0.299$
satisfy the system of inequalities above, with
$71.997984 < 71.998224 < 72 < 72.008064$ (angles in degrees)
for $\alpha$, $\theta_m$, $\tmean$, and $\beta$, respectively.

\begin{lemma}
\label{lem:realizable}
For each integer $n > 4$, there exists two real numbers $l_1$ and
$l_2$ satisfying simultaneously all the constraints of
Section~\ref{sec:construction} (i.e., Inequalities
\ref{ieq:l1-upper}, 
\ref{ieq:tmean-upper},
\ref{ieq:l2-upper}, 
\ref{ieq:l2-lower},
\ref{ieq:l1-plus-l2}, and
\ref{ieq:alpha-upper}).
\end{lemma}

In the sequel, we assume some choice of parameters has been made such
that all the constraints of this section hold.

\section{Restricted Configurations}
\label{sec:restricted-config}


Referring to Figure~\ref{fig:restricted-config}, recall from the
definition of $\beta$ (Equation~\ref{eqn:define-beta}) that $\theta <
\beta$ implies that circle \calc\ properly intersects link $OA'$.
Thus, points $P$ and $Q$ are well-defined.  When joint $B$ is on the
small arc $PQ$, recall that points $R$ and $S$ are also well-defined.
And finally, joint $C$ also fits inside $\Box ORBS$, due to
Inequality~\ref{ieq:l2-upper}.

\begin{smalldef}
A petal configuration is said to be {\em restricted} if the following
conditions hold:
\begin{enumerate}[(i)]
\item $\theta < \beta$,
\item $B$ is on the open small arc $PQ$ of \calc, and
\item $C$ is in the open region bounded by the 
        quadrilateral $\Box ORBS$.
\end{enumerate}
\end{smalldef}

Note in Figure~\ref{fig:restricted-config} that $\|BR\|$ and $\|BS\|$
are both smaller than $\|BC\|$.  We show now that this is always the case in a
restricted configuration.

\begin{lemma}
\label{lem:BR-BS-bound}
In a restricted configuration, both $\|BR\|$ and $\|BS\|$ are strictly
less than $l_2$.
\end{lemma}

\begin{proof}
For a point $X$, let $d(X,L)$ denote the 
distance from $X$ to the line $L$ through $O$ and $A'$.
We have that $\|BR\| = d(B,L)$.
Because $B$ is on the small arc $PQ$, $d(B,L) \leq d(Q,L) = \|OQ\|
\sin\theta$.  Hence,
\begin{equation}
\label{ieq:BR-upper}
        \|BR\| \leq \|OQ\| \sin\theta.
\end{equation}
By similar reasoning,
\begin{equation}
\label{ieq:BS-upper}
        \|BS\| \leq \|OP\| \sin\theta.
\end{equation}

Note that $\|OQ\| = 1 - l_1 = 1 - \sin\beta$, by
Equation~\ref{eqn:define-beta}.  For $x \in [0,\pi/2]$, $\sin x +
\cos x \geq 1$, so $\|OQ\| \leq \cos\beta$.  Because $l_P$ is an
increasing function, $\|OP\| = l_P(\theta) \leq l_P(\beta)$, and
$l_P(\beta) \leq \cos\beta$, by Equation~\ref{eqn:define-lP}.

Both $\|OP\|$ and $\|OQ\|$ are $\leq \cos\beta$.  Hence, by
Inequalities~\ref{ieq:BR-upper} and~\ref{ieq:BS-upper}, both $\|BR\|$
and $\|BS\|$ are bounded from above by $\cos\beta \sin\theta < \cos\beta
\sin\beta$, as $\theta < \beta \leq \pi/2$ by assumption.  But this is strictly
less than $l_2$ by Inequality~\ref{ieq:l2-lower}.
\end{proof}

\begin{lemma}
\label{lem:thetam-is-minimum}
In a restricted configuration, $\theta \ge \theta_m$.
\end{lemma}

\begin{proof}
Suppose, for contradiction, the petal is in a restricted configuration
with $\theta < \theta_m$.

Consider the two triangles $\triangle OAB$ and $\triangle OAP$.  These
triangles share the common side $OA$, and $\|AB\| = \|AP\| = l_1$.
Two sides of $\triangle OAB$ are equal length with two sides of
$\triangle OAP$.  Moreover the included angles satisfy $\angle OAB <
\angle OAP$, since by definition of restricted configuration
(condition (ii)), joint $B$ lies on the small arc $PQ$ of \calc.
Applying the cosine law to the remaining side in each triangle (or
using Euclid's Proposition 24, Book I) we see $\angle OAB < \angle
OAP$ implies $\|OB\| < \|OP\|$.  By definition
(Equation~\ref{eqn:define-lP}) $\|OP\| = l_P(\theta)$, which is less
than $l_P(\theta_m)$ since $l_P$ is an increasing function.  A direct
computation shows that $l_P(\theta_m) = l_2$, so we have $\|OB\| <
l_2$.

By Lemma~\ref{lem:BR-BS-bound}, we have also $\|BR\| < l_2$ and
$\|BS\| < l_2$.

All four points $O$, $R$, $B$ and $S$ are strictly inside the circle
of radius $l_2$ centred at $B$.  Joint $C$ is of course on this
circle, so it cannot be inside $\Box ORBS$.  This contradicts
condition (iii) of a restricted configuration.
\end{proof}

\begin{lemma}
\label{lem:remains-restricted}
Consider a petal in a restricted configuration.  Throughout any motion
during which $\theta$ is strictly less than $\beta$, the petal remains
in a restricted configuration.
\end{lemma}

\begin{proof}
Note that the points $P$, $Q$, $R$, $S$, and the circle $\calc$, are
defined by the positions of the joints $A$ and $B$; as the joints
move, so do the points $P$, $Q$, etc.  For simplicity, we omit
displaying this dependence on time.

Consider, in turn, the three conditions required for a restricted
configuration.  Condition (i) holds throughout the motion by assumption.

In any configuration, $B$ must be on \calc.  Since $B$ starts the
motion on the small arc $PQ$, for condition (ii) to be violated, $B$ must pass
through point $P$ or through point $Q$.  $P$ and $Q$ are on the
interior of links $OA'$ and $OA$, respectively, and 
since $\theta < \beta$, \calc\ always
properly intersects
these links.  Thus $B$ may not move through $P$ or $Q$, and hence condition
(ii) holds throughout the motion.

Given that $B$ remains on the small arc $PQ$, points $R$ and $S$ are well
defined.  As $C$ starts the motion inside $\Box ORBS$, condition
(iii) is violated only if $C$ passes through one of the sides of this
quadrilateral.  Sides $OR$ and $OS$ are portions of links, so $C$ may
not pass through them.  By Lemma~\ref{lem:BR-BS-bound}, $\|BR\|$
and $\|BS\|$ are both strictly less than $\|BC\|$, so $C$ cannot pass
through side $BR$ or side $BS$.  Thus, condition (iii) holds throughout
the motion.
\end{proof}

\section{Trees Can Lock}
\label{sec:trees-can-lock}

This section describes our main result: two inequivalent configurations
of a tree linkage.

Recall that $\theta_m$ is defined by a triangle with sides $1$, $l_1$,
and $l_2$, pictured in Figure~\ref{fig:thetam}.
With $\triangle OAB$, the cosine rule yields
\[
        l_1^2 = l_2^2 + 1 - 2l_2\cos\theta_m,
\]
while from Inequality~\ref{ieq:l2-upper} we can obtain
\[
        l_1^2 < l_2^2 + 1 - 2l_2\cos\tmean,
\]
which implies $\theta_m < \tmean$.  Putting this together with
Inequalities~\ref{ieq:alpha-upper} and
\ref{ieq:tmean-upper}, we have
\begin{equation}
\label{ieq:alpha-beta-range}
        \alpha < \theta_m < \tmean < \beta,
\end{equation}
hence $(\alpha, \beta)$ is a non-empty interval.  

Consider a tree in a configuration in which the petal configurations
are all congruent, with petal angles all equal to $\tmean \in (\alpha,\beta)$.  
We place joint $B$ on the small arc $PQ$.  
Because $l_2 < l_P(\tmean) = \|OP\|$ (Inequality~\ref{ieq:l2-upper})
we can place $B$ near, but not at, $P$ to ensure $\|OB\| > \|BC\|$.
This ensures that joint $C$ may be placed along $OB$, which is inside
$\Box ORBS$.  This is a valid configuration, and furthermore each
petal configuration restricted.  The next theorem shows all petals
remain bounded from below by $\alpha$.

\begin{theorem}
\label{thm:main}
Consider a tree of $n$ petals in a configuration such that $\theta_i
\in (\alpha,\beta)$ for $0 \le i < n$, and with each petal in a
restricted configuration.  During any motion, all petal
angles remain in the range $(\alpha,\beta)$.
\end{theorem}

\begin{proof}
Suppose, to the contrary, a motion exists that takes some petal angle
out of the range $(\alpha,\beta)$.  
Let $t_\alpha$ be the first instant that some petal angle, say for
petal $k$, reaches $\alpha$.  Let $t_\beta$ be the first instant that
some petal angle reaches $\beta$.

If $t_\beta < t_\alpha$, then at time $t_\beta$ all angles
are strictly greater than $\alpha$ and at least one is equal to
$\beta$.  This means
\[
        2\pi = \sum_{i = 0}^{n-1} \theta_i > (n-1)\alpha + \beta = 2\pi,
\]
a contradiction.  Hence $t_\alpha \le t_\beta$.

During the supposed motion, the joint angles change continuously
in time.  Since $\alpha < \theta_m$ by
Inequality~\ref{ieq:alpha-upper}, 
and $\theta_k$ approaches $\alpha$ from above as $t$
approaches $t_\alpha$ from below,
we may choose $t_0 < t_\alpha$ such
that $\theta_k \in (\alpha,\theta_m)$ at time $t_0$.

Note that during the motion up to time $t_0$, all petal angles are
strictly less than $\beta$, as $t_0 < t_\alpha \le t_\beta$.
By Lemma~\ref{lem:remains-restricted} all
petals remain in a restricted configuration before time $t_0$, so
Lemma~\ref{lem:thetam-is-minimum} applies to petal $k$.  This means
$\theta_k \ge \theta_m$, contradicting the choice of $t_0$.
\end{proof}

Recall that two simple configurations of a tree linkage are equivalent
if one can be moved to the other.  Our main result
is that a tree linkage can have two inequivalent configurations:
one with all petals in a restricted configuration, and the other with
one or more petal angles less than $\alpha$.  These configurations
are illustrated in Figure~\ref{fig:inequivalent}.  This result can 
easily be extended to the following corollary.

\begin{corollary}
There exist $N$-link tree linkages
such that the linkages have $2^{\Omega(N)}$ equivalence classes of
simple configurations.
\end{corollary}

  \begin{figure}[th]
  \begin{center}
  \input{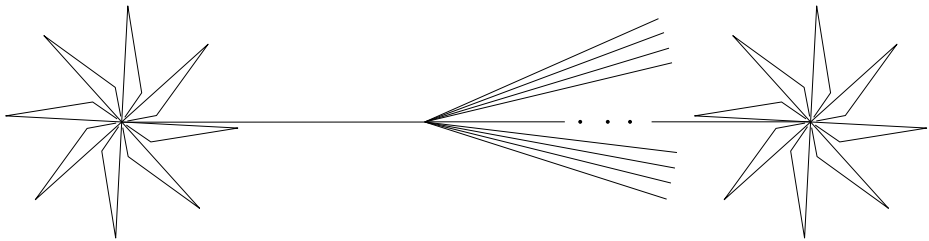}
  \caption{A tree linkage formed by joining $k$
copies of a lockable tree.  Each subtree may be in either an open
(as is the middle subtree) or a closed (the first and last subtrees)
configuration.  This linkage has at least $2^k$ configuration
classes.}
  \label{fig:exponential-example}
  \end{center}
  \end{figure}

\begin{proof}
Consider the linkage in Figure~\ref{fig:exponential-example}, in which
there are $k$ copies of an eight-petal lockable tree connected by long
links joining the $O$ joints of the subtrees.  The connecting links
are long enough that when they are stretched out to form a straight
chain, each subtree can be in either an open or a closed configuration
without crossing links.

Consider simple configurations in which the long links form a straight
chain, and each subtree is in either an open or a closed
configuration.  Label such configurations by a $k$-bit vector,
specifying for each subtree, whether its configuration is open or
closed.  Configurations with different labels are clearly not
equivalent, as a motion of the entire linkage that would take
some subtree in a closed configuration to an open configuration would
imply, by removing links outside the subtree, the existence of a
motion that would make petal angle of the subtree inferior to $\alpha$.  
Hence the number of
inequivalent configurations is at least $2^k \in 2^{\Omega(N)}$, as $N = 24k
+ (k-1)$.
\end{proof}

\section{Conclusion}
\label{sec:conclusion}

While no chain or polygon in the plane may lock, we showed in this report
that for a tree linkage can; i.e. that there can be
more than one equivalence class of simple configurations.  Indeed,
some $N$-link trees have $2^{\Omega(N)}$ equivalence classes.

\begin{figure}[th]
  \begin{center}
  \input{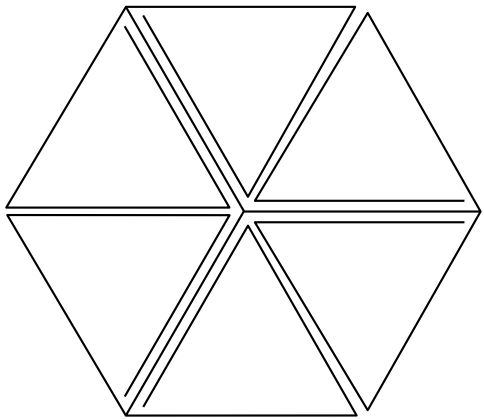}
  \caption{A nearly equilateral tree linkage
configuration.}
  \label{fig:near-equilateral}
  \end{center}
  \end{figure}

The tree construction of Section~\ref{sec:construction} constrains the
link lengths to be non-equal (it appears difficult to even get them
nearly equal).  This prompts the following question: can a tree
linkage with equal-length links have a locked configuration?  One
``nearly equilateral'' tree linkage is shown in
Figure~\ref{fig:near-equilateral}.  We conjecture that if the link
lengths are very nearly equal, this configuration is locked; the
intuition is as follows.  Each of the six triangular petals cannot
collapse, so each remains nearly an equilateral triangle.  If that is
the case, it seems that each petal can only move by pivoting about its
degree-3 joint, which it cannot do without crossing a link of the
adjacent petal.

Notice, however, that if the links truly are of equal length, the
configuration pictured cannot be simple.  

The nearly equilateral example has the feature that the graph has
maximum degree three.  It is easy to replace a high-degree joint with
a number of degree-three joints joined by a chain of tiny links.  Do
equilateral locked tree linkages with maximum degree three exist?

Finally, many interesting questions can be posed for linkages moving
in higher dimensions.  See~\cite{chains3d,chains4d} for recent work on
chain and cycle linkages moving under simple motion in three and more
dimensions.

\section*{Acknowledgments}
The research reported here was initiated at the 
International Workshop on Wrapping and 
Folding, co-organized by Anna Lubiw and 
Sue Whitesides, at the Bellairs Research 
Institute of McGill University, January 31 --
February 6, 1998.  We would like to thank Hazel Everett for many
useful discussions.

Our locked tree was inspired by a polygonal chain designed by Joe
Mitchell, kindly shared with one of the coauthors.  In particular, we
borrowed the circular structure from his example.

A preliminary version of this work appeared in~\cite{cccg-version}.

\appendix

\section{Proof of Lemma~\ref{lem:realizable}}
\label{app:realizable-proof}

This appendix proves the following lemma.

\begin{quote}
For each integer $n > 4$, there exists two real numbers $l_1$ and
$l_2$ satisfying simultaneously all the constraints of
Section~\ref{sec:construction} 
(i.e., Inequalities
\ref{ieq:l1-upper}, 
\ref{ieq:tmean-upper},
\ref{ieq:l2-upper}, 
\ref{ieq:l2-lower},
\ref{ieq:l1-plus-l2}, and
\ref{ieq:alpha-upper}).

\end{quote}

\begin{proof}
  
We show that $l_1 = \sin(\tmean + \epsilon)$ and $l_2 = l_P(\tmean -
\epsilon / n)$ are feasible link lengths, where $\epsilon =
0.01\deg$ is feasible for $n=5,n=6$ and any $0 < \epsilon < 0.4\deg$
is feasible for $n \ge 7$.  The proof consists of checking, in turn,
each of the constraints mentioned above.

\proofitem{Constraint \ref{ieq:l1-upper}: $l_1 < 1$}

Given our choice of $l_1$, this is satisfied as long as $\tmean +
\epsilon < \pi/2$.  Recall that $\tmean = 2\pi/n$ and $n \ge 5$
(Inequality~\ref{ieq:n-lower}), so $\tmean \le \frac{2\pi}{5}$.
Since $\epsilon < 0.4\deg$, the constraint is satisfied.

\proofitem{Constraint \ref{ieq:tmean-upper}: $\tmean < \beta$}

By the definition of $\beta$ (Equation~\ref{eqn:define-beta}),
and our choice of $l_1$,
we have $\beta = \tmean + \epsilon$.  
Since $\epsilon > 0$, the constraint is satisfied.

\proofitem{Constraint \ref{ieq:l2-upper}: $l_2 < l_P(\tmean)$}

Since $\epsilon > 0$, $\tmean - \epsilon/n < \tmean$.  Using the
definition $\tmean = 2\pi/n$, we see that $\tmean - \epsilon/n = (2\pi
- \epsilon)/n$.  Since $\epsilon < 0.4\deg$, $\tmean - \epsilon/n > 0$.
Now, $l_P(\theta)$ is an increasing function on $[0,\beta] \supset
(0,\tmean)$, and $l_2 = l_P(\tmean - \epsilon / n)$, so the constraint
is satisfied.

We summarize, for future reference, some inequalities derived so far,
\begin{equation}
\label{ieq:summary}
    0 < \tmean - \epsilon/n 
      < \tmean 
      < \beta = \tmean + \epsilon 
      < \pi/2.
\end{equation}

\proofitem{Constraint \ref{ieq:l2-lower}: $l_2 > \sin\beta \cos\beta$}

This constraint is the only one for which we distinguish cases based
on $n$.  For $n=5$ and $n=6$, a direct calculation shows that the
given link lengths (with $\epsilon = 0.01\deg$) satisfy the
constraint.

For the case $n \ge 7$, we show that any $0 < \epsilon < 0.4\deg$
yields $l_2 > \frac{1}{2}$, whence the constraint
follows since $\sin\beta\cos\beta = \frac{1}{2}\sin(2\beta) \le
\frac{1}{2}$ (the equality is an identity, the inequality uses 
$\sin x \le 1$).  

Plugging our choice of $l_2$ into the definition of $l_P$
(\ref{eqn:define-lP}),
\begin{equation}
\label{eqn:l2-lower-proof}
        l_2 = l_P(\tmean - \epsilon/n)
            = \cos(\tmean - \epsilon/n) 
              - \sqrt{l_1^2 -\sin^2(\tmean - \epsilon/n)}.
\end{equation}

Cosine is a decreasing function on $[0,\pi]$, so $\cos(\tmean -
\epsilon/n) > \cos\tmean > \cos(\frac{2\pi}{7})$ 
given $0 < \tmean - \epsilon/n < \tmean < 2\pi/7$
(\ref{ieq:summary} and $n \ge 7$).

Under the radical of~\ref{eqn:l2-lower-proof}, substituting 
$l_1 = \sin(\tmean + \epsilon)$ gives the expression
$\sin^2(\tmean + \epsilon) - \sin^2(\tmean - \epsilon/n)$.  Using the 
identity $\sin^2 x - \sin^2 y = \sin(x+y)\sin(x-y)$, this expression
becomes 
$\sin(2\tmean + \epsilon - \epsilon/n)\sin(\epsilon + \epsilon/n) 
\leq \sin(\epsilon + \epsilon/n) 
< \sin(2\epsilon)
< \sin(0.8\deg)$.
The last two inequalities follow by noting that 
$0 < \epsilon + \epsilon/n < 2\epsilon < 0.8\deg < \pi/2$, 
and sine is increasing on the interval $[0,\pi/2]$.

Using these bounds for the two terms in
Equation~\ref{eqn:l2-lower-proof},
\[
        l_2 > \cos\left(\frac{2\pi}{7}\right) - \sqrt{\sin(0.8\deg)}
           \approx 0.505 > \frac{1}{2}.
\]

\proofitem{Constraint \ref{ieq:l1-plus-l2}: $l_1 + l_2 > 1$}

Define function $f(\theta) = l_1 + l_P(\theta)$ on $[0,\beta]$.  We
know $l_P$ is a strictly increasing function on $[0,\beta]$, hence so
is $f$.  Furthermore, $f(0) = 1$, so $f(\theta) > 1$ for $\theta \in
(0,\beta]$.  

From \ref{ieq:summary}, we see $\tmean - \epsilon/n \in (0,\beta)$, so
$f(\tmean - \epsilon/n) > 1$.  By definition of $f$, this becomes $l_1
+ l_P(\tmean - \epsilon/n) > 1$.  Since $l_2 = l_P(\tmean -
\epsilon/n)$, we obtain that $ l_1 + l_2 > 1$ as desired.

\proofitem{Constraint \ref{ieq:alpha-upper}: $\alpha < \theta_m$}

By definition, $\alpha = (2\pi - \beta)/(n-1)$
(Equation~\ref{eqn:define-alpha}).  This shows $\alpha \ge 0$.
Rewriting $2\pi$ as $n\tmean$, and $\beta$ as $\tmean + \epsilon$
(\ref{ieq:summary}), the equation for $\alpha$ becomes $\alpha =
(n\tmean - \tmean - \epsilon)/(n-1) = \tmean - \epsilon/(n-1) < \tmean
- \epsilon/n$.  Combining all this with (\ref{ieq:summary}), we find
\begin{equation}
\label{ieq:alpha-summary}
    0 < \alpha < \tmean - \epsilon/n < \tmean < \beta < \pi/2.
\end{equation}

From (\ref{ieq:alpha-summary}) we see that
$\alpha \in (0,\pi)$.  By definition
(Equation~\ref{eqn:define-thetam}), $\theta_m$ is also in $(0,\pi)$.
Since cosine is decreasing on this interval, the constraint $\alpha <
\theta_m$ holds if, and only if, $\cos\alpha > \cos\theta_m$.  Using
the definition of $\theta_m$ (Equation~\ref{eqn:define-thetam}), this
latter inequality becomes $2 l_2 \cos\alpha > 1 - l_1^2 + l_2^2$.
We collect the terms in $l_2$ to one side, 
$l_2^2 - 2 l_2 \cos\alpha < l_1^2 - 1$, and complete the square
to get
\[
    (l_2 - \cos\alpha)^2 < l_1^2 - \sin^2\alpha.
\]
Since $l_1 = \sin\beta$ (by definition of $\beta$,
Equation~\ref{eqn:define-beta}), the right hand side can be written as
$\sin^2\beta - \sin^2\alpha$, which is positive since $\alpha < \beta$
by (\ref{ieq:alpha-summary}).  Since both
sides of the inequality are positive, we can take square roots which 
leads to
\[
    |l_2 - \cos\alpha| < \sqrt{l_1^2 - \sin^2\alpha},
\]
so
\[
      -\sqrt{l_1^2 - \sin^2\alpha} 
    < l_2 - \cos\alpha
    < \sqrt{l_1^2 - \sin^2\alpha},
\]
and we deduce that $l_2$ (which equals $l_P(\tmean - \epsilon/n)$) must satisfy
\begin{equation}
\label{ieq:l2-range}
        \cos\alpha - \sqrt{l_1^2 - \sin^2\alpha} 
    < l_P(\tmean - \epsilon/n) <  
        \cos\alpha + \sqrt{l_1^2 - \sin^2\alpha}.
\end{equation}

Comparing the left inequality of this with the definition of $l_P$
(\ref{eqn:define-lP}), we see that the former can be written
$l_P(\alpha) < l_P(\tmean - \epsilon/n)$.  From
(\ref{ieq:alpha-summary}) we note that $\alpha$ is less than $\tmean -
\epsilon/n$ and both quantities are in the range $[0,\beta]$.  The
increasing property of $l_P$ ensures that the left inequality of
(\ref{ieq:l2-range}) is satisfied.

For the upper bound, note that by (\ref{ieq:summary}), $\tmean -
\epsilon/n < \tmean$ and both quantities are in the range $[0,\beta]$.
The increasing property of $l_P$ ensures that $l_P(\tmean -
\epsilon/n) < l_P(\tmean)$.  By the definition of $l_P$
(\ref{eqn:define-lP}), $l_P(\tmean) < \cos\tmean$.  Given $0 < \alpha
< \tmean < \pi/2$ (\ref{ieq:alpha-summary}), the decreasing nature of
cosine on this interval ensures $\cos\tmean < \cos\alpha$.  Putting
this together, we find $l_P(\tmean - \epsilon/n) < \cos\alpha$, so the
upper bound of Inequality~\ref{ieq:l2-range} is satisfied.

\end{proof}

\bibliographystyle{abbrv}
\bibliography{linkages}

\end{document}